\documentclass[preprint,endfloats*,prb,aps,showpacs]{revtex4-1}

\usepackage[final]{graphicx}
\usepackage[dvipsnames]{color}
\usepackage{dcolumn}
\usepackage{hhline}
\usepackage{amsmath}
\DeclareGraphicsExtensions{.pdf,.eps}

\newcommand{\be}{\begin{equation}}
\newcommand{\ee}{\end{equation}}
\newcommand{\bea}{\begin{eqnarray}}
\newcommand{\eea}{\end{eqnarray}}

\newcommand{\isen}{Institute for Electronics, Microelectronics, and Nanotechnology (IEMN), CNRS UMR 8520, Dept. ISEN, 59652 Villeneuve d'Ascq Cedex, France}
\newcommand{\luxembourg}{Physics and Material Sciences Research Unit,
University of Luxembourg, 162a ave. de la Fa\"iencerie,
L-1511 Luxembourg}
\newcommand{\grenoble}{Institut N\'eel,
CNRS, 25 rue des Martyrs BP 166, 38042 Grenoble
cedex 9 France} 
\newcommand{\rome}{Istituto di Struttura della Materia (ISM), Consiglio Nazionale delle Ricerche, Via Salaria Km 29.5, CP 10, 00016 Monterotondo Stazione, Italy}
\newcommand{\sanseb}{Nano-Bio Spectroscopy Group and ETSF Scientific Development Centre, Departamento de F\'isica de Materiales, Centro de F\'isica de Materiales CSIC-UPV/EHU-MPC and DIPC,  Universidad del Pa\'is Vasco UPV/EHU, Av. Tolosa 72, E-20018 San Sebastián, Spain}
\newcommand{\fhi}{Fritz Haber Institut der Max Planck Gesellschaft, Faradayweg 4-6, 14195, Berlin,Germany}

\begin{document}

\title{Efficient Gate-tunable light-emitting device made of defective boron nitride nanotubes: from ultraviolet to the visible}

\author{Claudio Attaccalite${^1}$,  Ludger Wirtz${^{2,3}}$, Andrea Marini${^4}$, Angel Rubio${^{5,6}}$\footnote{To whom correspondence
should be addressed; email: angel.rubio@ehu.es}}

\affiliation{
$^1$ \grenoble \\
$^2$ \isen \\
$^3$ \luxembourg\\
$^4$ \rome\\
$^5$ \sanseb\\
$^6$ \fhi}

\date{\today}

\maketitle



{\bf Boron nitride is a promising material for nanotechnology applications due to its two-dimensional graphene-like insulating and highly-resistant
structure.\cite{Pakdel2012256,wirtz_review,RevModPhys.82.1843} 
Recently it has received a lot of attention as a substrate to grow and isolate graphene\cite{bngraphene}  as well as for its intrinsic UV lasing response.\cite{watanabe1,watanabe2}  Similar to carbon, one-dimensional boron nitride nanotubes (BNNTs) have been theoretically predicted\cite{rubiobn} and later synthesised.\cite{tubesynthesis} 
 Here we use first principles simulations to unambiguously demonstrate that i) BN nanotubes inherit the highly efficient UV luminescence of hexagonal BN; ii) the application of an external perpendicular field closes the electronic gap keeping the UV lasing with lower yield; iii) defects in BNNTS are responsible for tunable light emission from the UV to the visible controlled by an transverse electric field (TEF). 
Our present findings  pave the road towards optoelectronic applications of BN-nanotube-based devices that are simple to implement
 because they do not require any special doping or complex growth. 
}           
\newpage

Scientists have worked hard in the last decades to grow defect free nano-structures. The near-perfect atomic arrangement at the nano-scale has been employed to create new efficient devices as light-emitters, transistors  and sensors. While many applications benefit from using defect free materials, the presence of particular impurities can generate new and fascinating properties. For example F-centre in ionic crystals have been widely used in luminescence applications.\cite{RevModPhys.18.384}  More recently nitrogen vacancies in diamond have been proposed for quantum computation.\cite{PhysRevA.71.060310}
Defects play an important role also for the luminescence properties of
hexagonal boron nitride and related nanostructures.\cite{PhysRevLett.100.189701,attaccalite2011,PhysRevB.75.085205,PhysRevB.77.235422,PhysRevB.78.155204}
\begin{figure*}[ht]
 \centering
 \includegraphics[width=1.\textwidth]{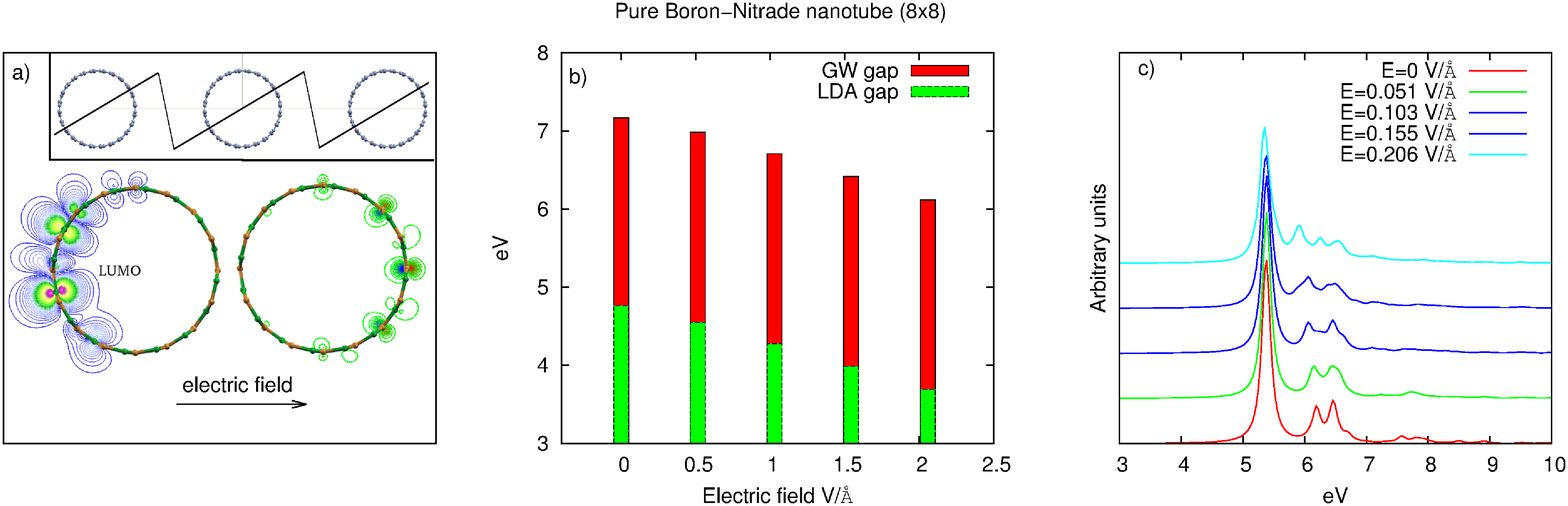}
\caption{\label{optics_pure} Panel $a)$ left(right): lowest conduction band (highest valence band) in presence of an external field E=0.206~V/{\AA}.  In panel $b)$ we report the band gap of BN $8\times8$ nanotube versus the TEF in LDA and $G_0W_0$ approximations; in panel $c)$ the corresponding optical absorption computed at the many-body Bethe-Salpeter level.}
\end{figure*}
Similarly to graphene, a single BN layer can be rolled up to form new structures ranging from single and multi-wall nanotubes to
BN-fullerens\cite{wirtz_review}. In contrast to graphite, the ionic character of the BN bond results in a wide band-gap of about 6 eV for bulk hexagonal-BN\cite{PhysRevLett.96.126104,PhysRevLett.100.189701,arnoud2006}. The combination of a such large gap with a strong electron-hole attraction makes the optical properties of hexagonal-BN based nanostructures largely independent of the layer arrangement and dimensionality.\cite{PhysRevLett.96.126104,wirtz_review} Although bulk h-BN has been shown to exhibit a strong luminescence, it cannot be used for optical applications in the visible range because  the emission frequency 
is fixed to about 5.75 eV in the UV.
However the presence of impurities can drastically modify this scenario, as it has been shown, theoretically\cite{attaccalite2011} and  experimentally.\cite{PhysRevB.75.085205,PhysRevB.78.155204,PhysRevB.77.235422}. \\ 
The defects electronic structure and formation energies have been widely studied in bulk h-BN and BN nanotubes.\cite{PhysRevB.67.113407,p911623m,jp056941l,berzina,jp800096s,zobelli} Luminescence in the visible was attributed to the presence of deep levels in the sample,\cite{ADMA200700493,nl0726151,Chen2010S353} whereas the UV emission is an intrinsic response (Frenkel-exciton) of the structurally perfect hexagonal BN.\cite{pierret2013excitonic}
Here we propose to engineer BNNTs with particular defects in order to generate light-emission in a wide range of frequencies, that can be tuned by means of an external electric field. The range of tunability of the proposed nanotube-based light emitting device depends from defect location and type. 

For pure BNNTs, it has been shown that the application of a transverse electric field generates a Stark effect leading to a strong reduction of the band gap\cite{PhysRevB.69.201401,PhysRevLett.94.056804}. 
The external field leads to a localisation of the conduction-band minimum/valence-band maximum
on opposite sides of the tube(see Fig.~\ref{optics_pure}). The corresponding energy shift of the band edges is thus proportional
to the nanotube diameter. Experimentally as-grown nanotubes contain defects that lead to both deep or shallow levels in the gap. The wave-functions of these levels are, a priori, only slightly affected by the presence of an external electric field\cite{attaccalite2007,jp800096s} because they are associated with localised orbitals centred on the impurity. However, their level position with respect to the bands edges changes\cite{jp800096s,attaccalite2007} because valence(conduction) bands are modified by the external field. Here we will show that this property can be employed to produce tunable and highly efficient bright light-emission devices based on defective BN nanotubes.

\section*{Results}
\begin{figure} 
 \centering
  \includegraphics[width=0.9\textwidth]{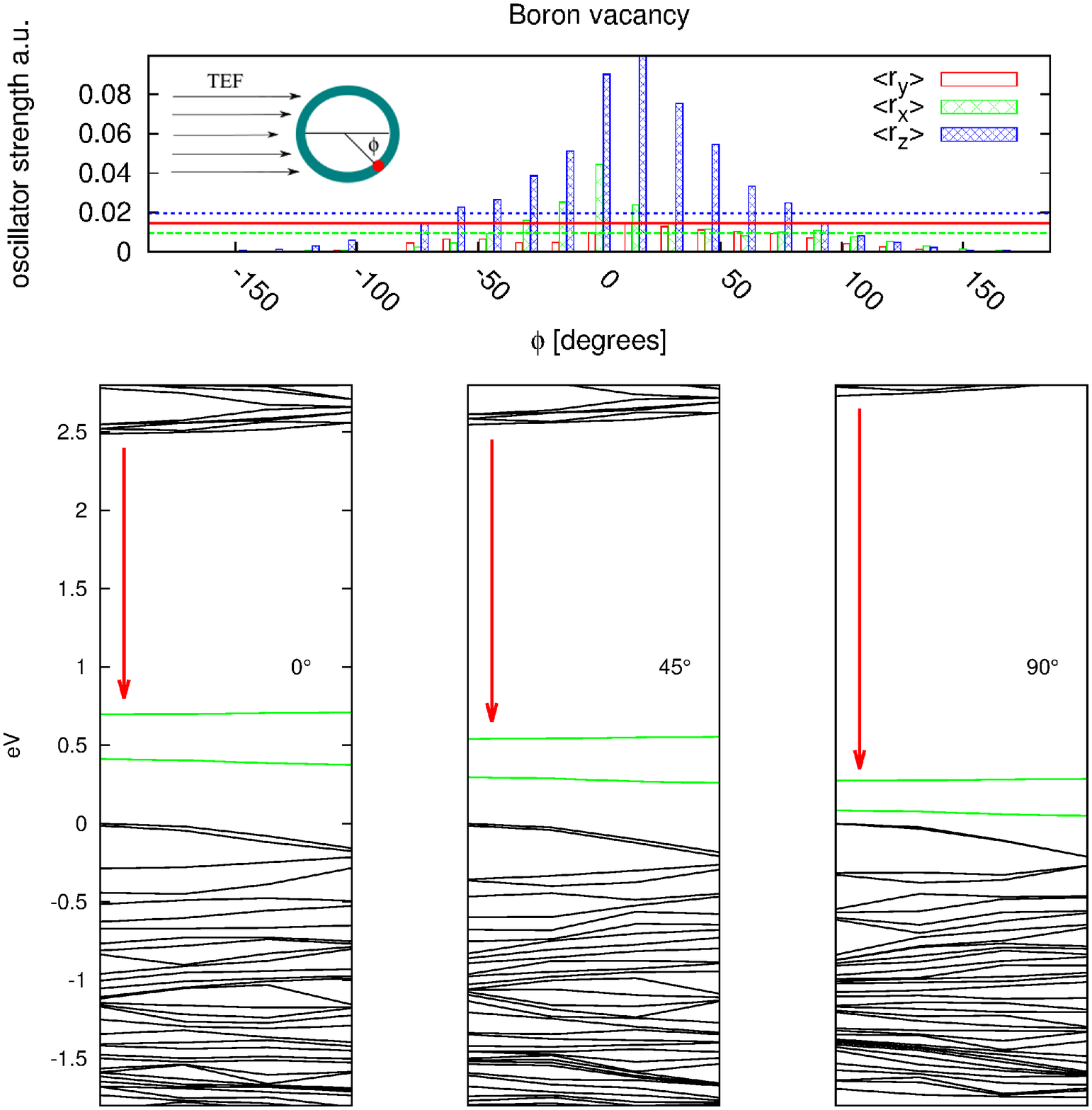}
 \caption{In the top panel we report the oscillator strength versus the angle between the defect and the bottom conduction band for $V_B$.  The optical matrix elements are averaged on the first conduction bands within an energy range of $0.15$~eV. The straight lines are the dipole matrix elements  at zero TEF. In the same figure it is present also a schematic representation of the a BN tube with a defect in presence of a TEF.  
 In the bottom panels we show the band structure of the same tube\label{bands_bvac} versus the angle between the defect and the electric field. The red arrow represents the transition responsible for the luminescence in presence of $V_B$.\cite{attaccalite2011} The intensity of the TEF is 0.206 V/{\AA}.}
\end{figure}

We model the electronic and structural properties of the pure and defected  BN nanotubes under a TEF using state-of-the-art first-principles methods based on density functional theory(DFT) combined with many-body perturbation-theory(MBPT) approaches. These methods allow calculation of quasiparticle-band structure and optical properties with a  high degree of accuracy (see Methods section for details). In the past this theoretical framework has been shown to be very efficient in predicting the electronic properties on BN nanotubes that were later on confirmed in the experiments.\cite{wirtz_review,rubiobn,PhysRevLett.96.126104,RevModPhys.82.1843}\\
We start our study by analysing the case of pure isolated BN nanotubes immersed in a static transverse electric field. A transverse electric field reduces the band gap, 
as shown in panel $b)$ of Fig.~\ref{optics_pure}. 
The gap reduction, induced by the TEF, is directly proportional to the electric field strength and to the tube diameter.\cite{PhysRevB.78.085423} Surprisingly the shrinking of the band-gap  slightly modifies the optical response of the tube.\cite{attaccalite2007} The main exciton remains in the same position, while a small fraction of
its spectral weight is redistributed to higher excitons (see panel $c)$ in Fig.~\ref{optics_pure}). In fact the conduction and
valence orbitals contributing to the gap reduction are localised on the opposite side of the tube, they have a very little overlap (see panel $a)$ of
Fig.~\ref{optics_pure}) and therefore their contribution to the optical response is negligible. Therefore we conclude that while the giant Stark effect present in BNNTs can modify their transport properties\cite{transport}, it leaves the optical response mainly unchanged.  The light emission spectra of \textit{pure} BNNTs is thus not tunable by an external electric field.
The presence of defects drastically modifies this picture.
Different experiments have shown that impurities induce light emission below 5 eV  in BN nanostructures,\cite{PhysRevB.75.085205,PhysRevB.78.155204,PhysRevB.77.235422} and modify the luminescence arising from the main bulk exciton.\cite{attaccalite2011}
These effects can be explained by the presence of deep levels in the BN band gap.\cite{attaccalite2011} The low frequency emissions are due to transitions from and to these levels. Moreover when the impurity levels are close to the top valence band or bottom conduction band they mix with the bulk excitations giving rise to a splitting of the main excitonic peak.\cite{attaccalite2011} 
Among the different impurities responsible for light emission we can distinguish two families: defect complexes and single defect centres. The first
family is formed by multiple defects as for instance di-vacancies, defect lines and so on. The second family consists in a single defect centre as
for instance boron (nitrogen) vacancies or a substitution of a boron(nitrogen) atom with a carbon one. The main difference between these two families lies in the different kind of transitions involved in light absorption/emission processes. In the case of defect complexes both donor and acceptor states are present in the band gap  while in the other case there is only a single donor or acceptor state. Therefore in the case of defect complexes the optical response is dictated by the quasi-donor-acceptor transitions\cite{Museur2008}, while in the simple defect centres light absorption/emission is due to transition between bulk states and deep defect levels (see Fig.~\ref{bands_bvac}). 

\begin{figure}[h]
 \centering
 \begin{tabular}{cc}
 \includegraphics[width=0.9\textwidth]{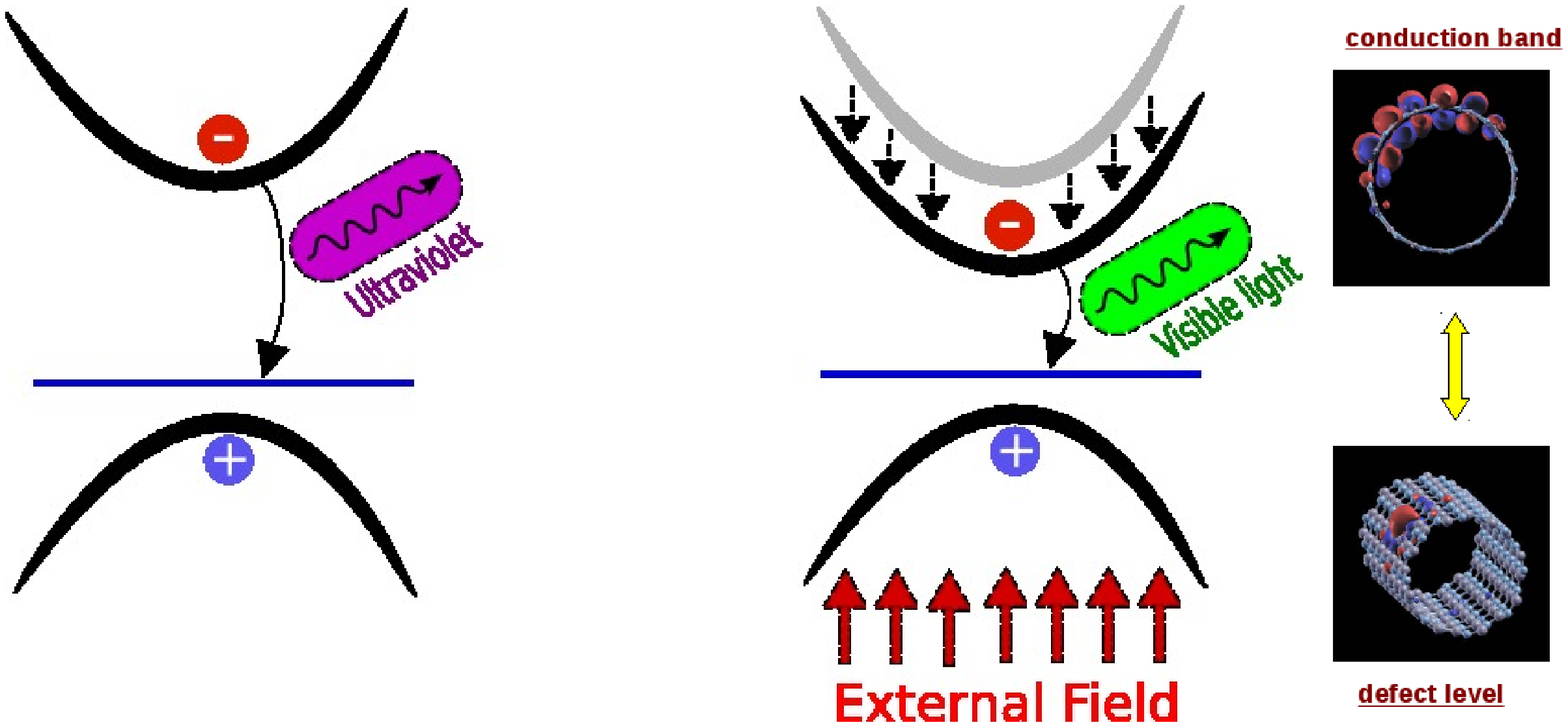} 
 \end{tabular}
\caption{Schematic representation of light emission process \label{schematic} for an acceptor impurity in a BN nanotube. On the left a simplified band structure picture in presence of a TEF. On the right conduction and defect orbitals responsible for the emission process.}
\end{figure}

The electronic structure of defects in BN nanotubes is similar to the one of defects in a single BN-sheet.\cite{PhysRevB.67.113407} 
In fact due to the large band gap curvature effects play a minor role  on the optical properties of pure BNNTs, where the strong localisation of excitons renders the optical spectra almost independent from the tube diameter and chirality.\cite{louietube,PhysRevLett.96.126104} In order to simulate a tube with defects we use the same methodology employed for the pure tube but with larger supercells in such a way to reduce the defect-defect interaction. Although large part of the tubes produced in the experiments are multi-wall and possess a zig-zag chirality, we chose a $12\times 12$ armchair one as prototype for our study. This choice is motivated by two reasons: first the primitive cell of an armchair tube, radius being equal, contains less atoms than a chiral one, second we expect only small differences with respect to the optical response of multi-wall or chiral nanotubes for the reasons discussed above.
When we turn on a TEF, the band gap of the tube shrinks and consequently, the  defect levels change position with respect to the band edges.\cite{jp800096s,attaccalite2007} The orbitals associated with defects levels are strongly localised on the impurities (see right panel of Fig.~\ref{schematic}) and therefore they are slightly deformed by the presence of the external field. 
To first order, the shift of the defects levels is thus given by the 
potential generated by the TEF and depends therefore on the position
of the defect respect to the direction of the electric field (see inset in Fig.~\ref{bands_bvac}).
This is visualised in the bottom panel of Fig.~\ref{bands_bvac} for three different defect positions.  We will show in the following, how this property gives rise to a tunable and efficient light emission.\\
In order to predict the emission frequencies of BN nanotubes in presence of 
defects, we used a simplified approach. The first necessary ingredient to get 
light emission is  non-vanishing optical matrix elements between 
the discrete donor(acceptor) state and the continuum states of the 
bottom conduction(top valence) bands. In the upper panel of Fig.~\ref{bands_bvac} we show
the strength of the optical matrix elements between the defect level and 
the bottom conduction bands for the case of a Boron vacancy, $V_B$. 
The optical matrix element displays a strong dependence on the angle between
the defect position and the electric field (see inset in the top panel of Fig.~\ref{bands_bvac}). The same phenomena occurs for other simple acceptor or donor defects like substitution of a Nitrogen or a Boron atom by a carbon one, $C_B$ and $C_N$ respectively.
Furthermore, we note that the optical matrix elements for polarisation
along the tube axis (z-axis) dominate, which also holds
for the optical response of pure nanotubes\cite{wirtz_review}.\\
The presence of the external electric field localises valence and conduction bands on opposite sides of the tube (see Fig.~\ref{optics_pure}), therefore  transitions are maximal only when the defect is aligned with the 
bottom(top) of the conduction(valence) band. The dipole element decreases
to almost zero as the defect is turned to the opposite side of the tube.
Consequently one can expect that luminescence, which is generated by 
transitions from and to the defects levels, will be efficient only when 
the defects are positioned on the side of "localised" conduction(valence) 
band edge.  This focusing effect increases with the tube size and field intensity. \\
Now that we are sure that transitions from and to simple defect centres are not zero in presence of a transverse electric field, we investigate
how the field modifies the emission frequencies.
In order to predict light emission we start from the quasi-particle(QP) band structure in presence of defects. We consider the energy differences for transitions between defect states and the top valence(bottom conduction) states. This allows us to investigate light-emission  versus transverse electric field, without including electron-hole interaction or lattice relaxation (see Methods section).  We found that also in presence of defects the GW renormalisation for conduction(valence) bands and defects levels is almost a constant respect to the external electric field (see also Fig.~\ref{optics_pure}(b)).

Luminescence can be estimated from the QP band structure in presence of TEF as the sum of independent transitions between conduction bands and defect states.
However light emission originates from  electron-hole recombination, a two-particle process that cannot be described by means of the quasi-particle band structure only. In fact electrons and holes attract each other and this attraction modifies the transition energies. These processes  can be naturally treated 
within a two-particle Green's functions formalism\cite{strinati}  and it has been shown that transition energies from and to defect states are strongly
renormalised\cite{attaccalite2011} by the electron-electron correlation. In order to model this correction, we calculate the exchange and electron-hole attraction between the defect level and the bottom conduction(top valence) bands only. In the past this approximation has been successfully employed to predict excitation energies of F-colour centres.\cite{louie}\\ 
In addition to the corrections originating from the electronic correlation, we have to consider the contributions due to the lattice relaxation
induced by the excited carriers. These are the so called Stokes and anti-Stokes shifts. The Stokes shifts can be estimated by means of a constrained DFT calculation with different defect occupation. 
We investigated three different defects, an acceptor the Boron vacancy $V_B$, a donor the Carbon substitution of a Boron atom $C_B$, and the  Boron-Nitrogen di-vacancy $V_{BN}$.
In order to estimate the Stokes shift we considered the case of a completely empty acceptor state(or a completely filled donor state). In principle one should consider also the correction coming from the partial filling (emptying) of the conduction(valence) bands, but this is supposed to be a minor effect because these bands are delocalised along the $\vec z$ direction.  In this way we obtain a rough estimation of the Stokes shift of 
$\Delta E_s \simeq 0.19~eV$ for $C_B$ and    $\Delta E_s \simeq 0.03~eV$ for $V_B$. We did not calculated any Stokes shift for $V_{BN}$  because in this case it is irrelevant as it will be clear in the following.
Vested with this theoretical approach we proceed in the study of light emission versus the external electric field.

\begin{figure}[ht]
 \centering
\includegraphics[width=1.\textwidth]{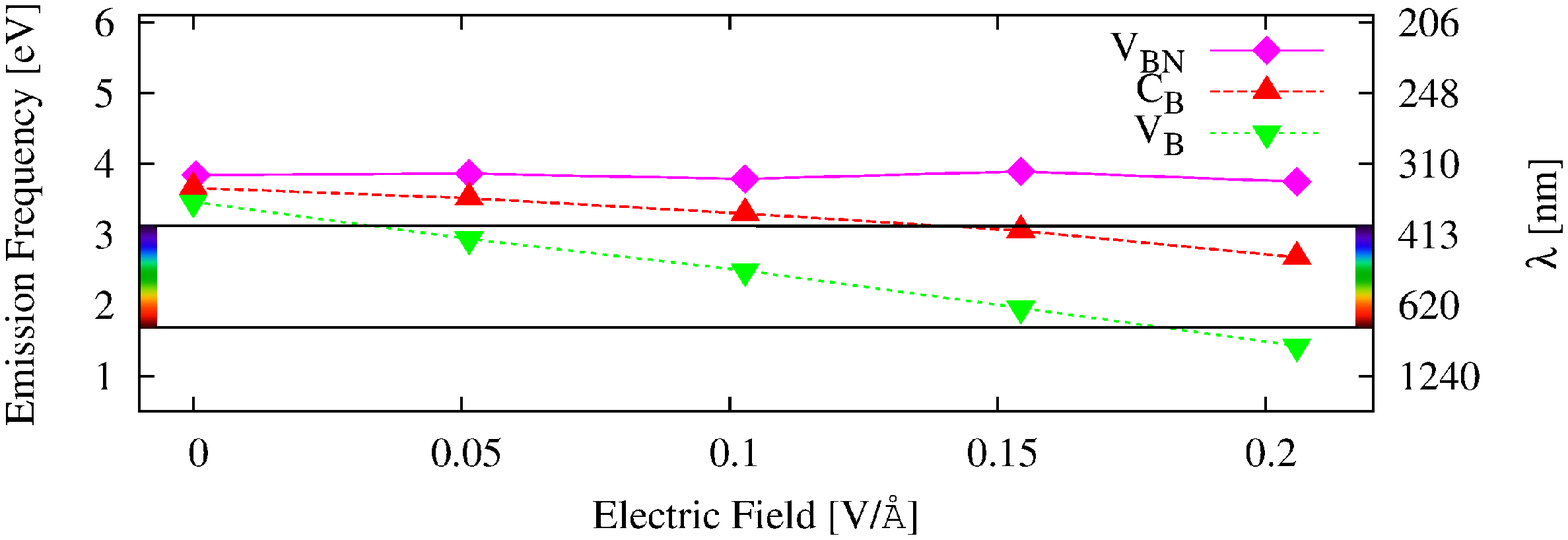}
\caption{Predicted light emission for different defects as a function of the transverse electric field. All the defects are taken in the position of maximum emission, according to their optical matrix elements, see also Fig.~\ref{bands_bvac}. In the  $V_{BN}$ we did not include any Stokes shift.\label{emission8}}
\end{figure}

\begin{figure}[ht]
\centering
 \includegraphics[width=1.\textwidth]{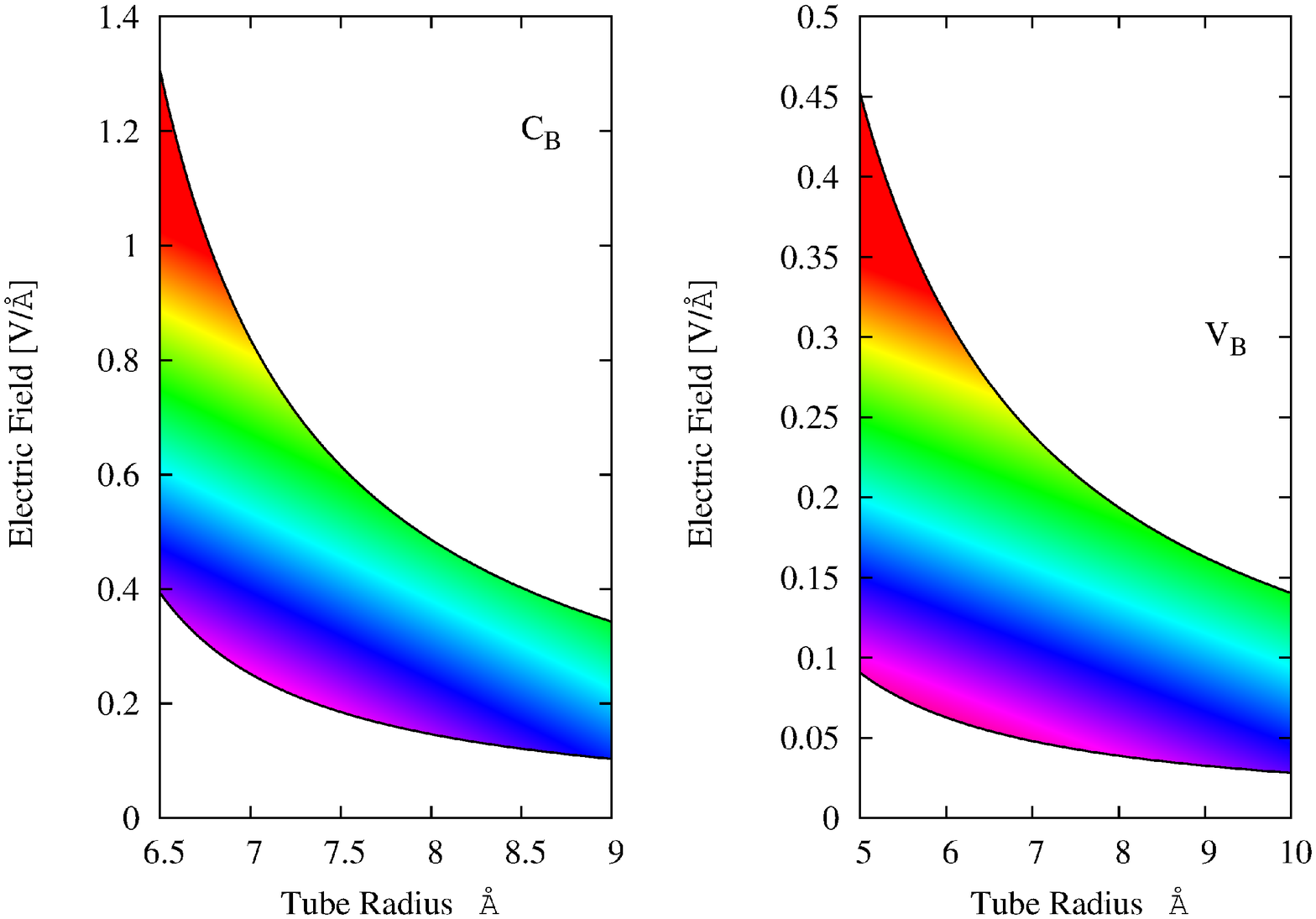}
\caption{Visible light emission range\label{emission} as function of the tube radius and the external electric field for $V_B$ and $C_B$ cases (rainbow colours are just a guide to the eyes).}
\end{figure}

We report our predicted light emission for $BN(12,12)$ tube versus the transverse electric field in Fig.~\ref{emission8}.   As one can see from the figure
an external electric field allows to vary the emission frequency in a large spectral range for the C$_B$ and V$_B$ cases. 
Notice that in presence of defect complexes, as for instance the BN di-vacancy $V_{BN}$, the emission frequency does not change with the external field. In fact in this case the emission is dominated by transitions between donor and acceptor states in the band-gap.\cite{Museur2008,attaccalite2011} Since the wave-functions associated to these states are localised on the impurity, the effect of external electric field is irrelevant. In the left panel of Fig.~\ref{schematic} we present a schematic representation of the light emission process 
from BN nanotubes in presence of defects. We want to underline that this process happens only when the defect is aligned with the conduction(valence) maximum, otherwise the emission will be inefficient due to the small dipole matrix elements.\\
Although the results of Fig.~\ref{emission8} can be theoretically extended to larger tubes, calculations become soon prohibitive due to the large
number of atoms, the vacuum in the super-cell and the number of conduction bands that enter in the many-body operators. Therefore in order to
predict light emission in larger (more realistic) tubes we assume many-body corrections to be a constant with respect to the tube size and we fit the emission energy with a simple  linear curve
\be
E_{emission} = E_0 + \alpha \xi 
\ee
where $\xi$ is the external electric field. This relation was already employed  to describe the band gap closing of h-BN nanotubes under the effect of a TEF in simple tight-binding models and \emph{ab-initio} calculations.\cite{PhysRevB.78.085423,Chegel2012154}
In principle the linear coefficient $\alpha$ depends on the tube size. In order to estimate this
dependence we performed different calculations at the DFT level, varying the tube size. We found that $\alpha$ changes
linearly with the tube radius $R$,  $\alpha(R) = \alpha_0 + R \beta$. A similar behaviour has been found for the pure BNNTs gap versus the electric-field and tube radius.\cite{PhysRevB.78.085423} Combining the  previous two equations we can predict the electric field $\xi$ necessary to produce light emission at a given frequency $E_1$:
\be
\xi = \frac{E_1 - E_0 }{\alpha_0 + R \beta}.
\label{eqfield}
\ee
Now we use Eq.~\ref{eqfield} to estimate the intensity of the TEF that will induce emission in the visible range ($1.65-3.1$~eV). In Fig.~\ref{emission} we report the visible emission range for the $V_B$ and $C_B$ cases versus the TEF intensity and tube radius. 

\begin{figure}[t]
\centering
 \includegraphics[width=1.\textwidth]{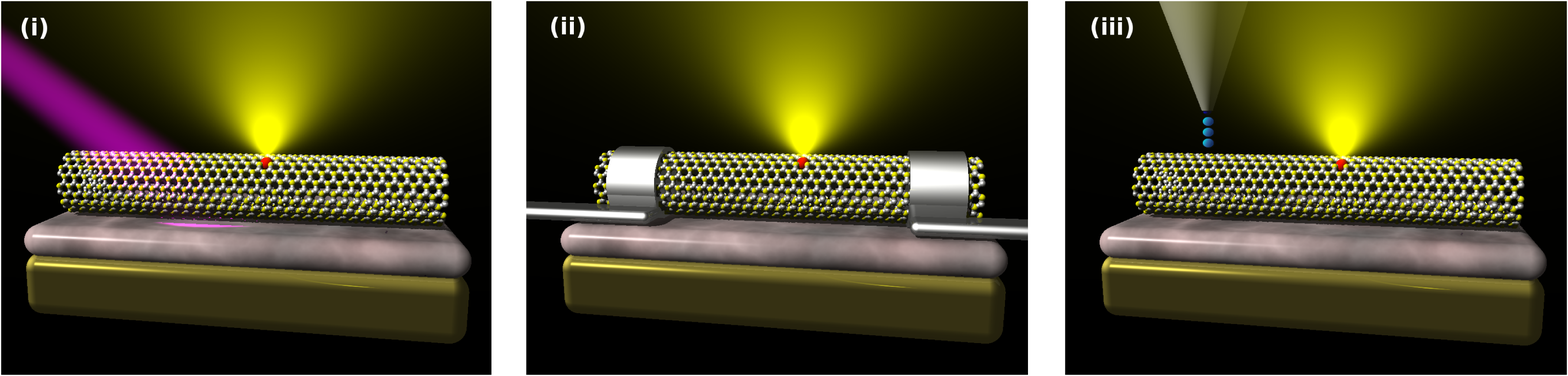}
\caption{\label{device} Schematic set-up for the suggested three possibilities to activate the optoelectronic device based on defective BN nanotubes (i) light-induced luminescence (ii) ambipolar transistor configuration (iii) 
electron induced electron-hole pairs by means of STM tip.}
\end{figure}

In general an increase of the tube size reduces the strength of the transverse electric field necessary to obtain emission in the visible range, For sufficient large tubes the TEF intensity is of the same order of the one available in small devices. 
Notice that an TEF produces an electrostatic potential inside the tube
that is proportional to the TEF intensity and the tube radius.\cite{jp800096s} Therefore Eq.~\ref{eqfield} breaks down for too large nanotubes or too strong fields.  However a giant Stark effects has been experimentally measured in pure BN nanotubes with a radius of about $23~\AA$ and a TEF of $0.08~V/\AA$. In the same experiment a gap reduction of more than $1~eV$ has been obtained.\cite{PhysRevLett.94.056804} Comparing these values with our extrapolation in Fig.~\ref{emission} it is clear that there is a large margin to produce visible light with experimentally accessible nanotubes and electric fields.\\
Finally we consider defects formation and their charge state. Recent experiments\cite{jin2009fabrication} have shown that it is possible to introduce defects in h-BN structures by means of electron irradiation. This process is mainly dominated by boron mono-vacancies even if other larger vacancies are present.  These vacancies can also be transformed in substitutional defects by introducing C atoms in the experiment,\cite{krivanek2010atom,risto}  and the final process can be controlled by charging the system during the irradiation.\cite{risto} These advances make possible the realisation of the device that we are going to discuss in the following. Regarding the charge state of the defects, in the present paper we investigated only neutral ones. Charged defects posses different relaxation energies and electronic structure. This fact influences also their optical properties, as it has been recently  shown in the case of vacancies in SiC.\cite{bockstedte2010many} The present results can easily be extended to charged defects and we expect that the main findings will remain valid. 
In fact the tunability of the light emission it is related to the localisation of defect states versus the delocalised bulk ones. Therefore a different charge state will modify the emission at zero field but not its behaviour in presence of a TEF.\\
Now that we have shown how to produce tunable light emission with defective BN nanotubes and discussed the feasibility of our idea, we briefly present the possible configurations of a device based on BN nanotubes. The generic configuration of the device (see Fig.~\ref{device}) comprises depositing as-grown BN nanotubes on an insulating surface (for example silicon oxide) acting as a dielectric to enable the application of the gated electric field that controls the light emission. The configurations is very much similar to the one of a field effect transistor (FET). The activation of the BN-defected optoelectronic device could be done by one of the following three processes: i) using UV light, ii) introducing an ambipolar 
current that recombines in the defect and emits light dictated by the applied 
gate voltage\cite{Chen18112005} iii) using tunnelling current through
an STM tip close to the nanotube. The excited electrons would inelastically 
decay very fast into the lowest energy state (the defect-liked Frenkel exciton) that would further decay by emitting light, again with a frequency dictated by the applied voltage, a process similar to the one leading to light emission in electronically excited semiconductors and fluorescent materials. A schematic set-up of those devices is illustrated in Fig.~\ref{device}.

\section*{Discussion}
In conclusion, we have shown that light emission from BNNTs with simple defect centres can be tuned by the presence of TEF. This opens the possibility to use these systems as light emitting devices. The use of (non-tunable) UV-light emitting devices based on crystalline hexagonal BN has already been  suggested before\cite{patent}. Here, we move one important step further by showing how nanotube based devices could operate in the UV and visible range by varying the external field. The external electric field, necessary to tune the emission,  can be applied using a field-effect transistor configuration\cite{transport}. The intensity necessary to produce visible light decreases with the tube size. The present results have been patented by some of the authors.\cite{tubepatent} 

Finally we envision that the present findings can be applied to other two-dimensional semiconducting or insulating layered materials that form tubes, as it is the case for transition metal dichalcogenides.\cite{cogenides} 
\section*{Methods}
{\small 
BNNTs are simulated by using a supercell approach,\cite{rubiobn} where
the tube is oriented along the $z$ direction, and a large empty space is left in the other two directions between tube replica in order to reduce the tube-tube interactions. Subsequently a sawtooth electric field (see inset in Fig.~\ref{optics_pure}(a)) with the cell periodicity is added along the $x$ direction.
In order to simulate light emission in BNNTs we employed a combination of Density-Functional Theory (DFT) plus Many Body Perturbation Theory (MBPT).
DFT is an exact theory for ground state properties and it is known to describe very well the structural properties of boron-nitride nanostructures  within Local Density Approximation (LDA). All DFT calculations have been performed using a $1\times1\times5$ supercell containing 240 atoms. The distance between the tube replica was 29 a.u.~, we used  a $1\times1\times2$ k-point sampling, LDA for the exchange correlation functional\cite{ceperley}, a plane waves cutoff of 45 Ry for the wave-function and norm-conserving pseudo-potentials.\cite{troullier} All DFT calculations have been performed with the PWSCF code\cite{pwscf} and the atomic structures have been relaxed using a BFGS quasi-Newton algorithm. 
Excited state and optical properties have been studied by means of MBPT. 
We calculated quasi-particle properties solving a Dyson equation within the so-called $G_0W_0$ approximation\cite{aryasetiawan1998gw,gw-strinati}, where all the Green's functions and the self-energy operator are constructed  with eigenvalues and eigenvectors of the Kohn-Sham(KS) Hamiltonian. 
Non-self consistent GW calculations have been performed with the code YAMBO\cite{yambo} using a plasmon pole approximation for the dielectric constant.  We used 30.000 G-vectors for the wave-function, 2 Ha for the response block size and 3000 bands for the screening. A cylindrical cutoff has been applied to the  Coulomb potential in order to reduce the tube-tube interaction. 
Neutral excitations, responsible for the absorption spectra were obtained from a two-particle Green's function equation, the Bethe-Salpeter equation, that is solved in the static ladder approximation\cite{strinati}, including excitonic effects.
We excluded quasi-free electron states\cite{PhysRevB.69.201401} in the Bethe-Salpeter equation, because they are not supposed to be responsible for luminescence.
We performed all calculations without including spin-polarisation effects. Even if we know that exchange-splitting slightly modifies the defects levels positions\cite{attaccalite2011,PhysRevB.76.014405}, this effect does not modify the main results of the paper.
For the large tubes employed to get the results in Fig.~\ref{emission8} we estimated the GW and electron-hole interaction from the one of a BN-sheet with the same defects and a distance between the periodic replica equal to the  inter-tube distance.\cite{attaccalite2011}}
\section*{Acknowledgements}
We acknowledge financial support from the European Research Council Advanced Grant DYNamo (ERC-2010-AdG - 267374) Spanish Grant FIS2011-65702-C02-01, Grupo Consolidado UPV/EHU del Gobierno Vasco (IT-319-07) and European Commission project CRONOS (280879-2).  Computational time was granted by i2basque and BSC "Red Espanola de Supercomputacion" and GENCI-IDRIS (Nos. 100063 and  No. 091827).  A. M. acknowledges funding by MIUR FIRB Grant No. RBFR12SW0J.
\section*{Author Contributions} CA, LW, AM and AR contributed to the discussions, theoretical analysis and writing of the manuscript. CA performed the calculations and AR designed the research.
\section*{Author Information}  The authors declare no competing financial interest. Correspondence and request or materials should be addressed to AR (angel.rubio@ehu.es).

\addcontentsline{toc}{chapter}{Bibliography}
\bibliographystyle{naturemag}

\end{document}